\input epsf.tex

\magnification\magstep 1
\font\ff = cmr9            % footnote font
\font\sy = cmsy10          % math symbols
\font\ref = cmr12          % References
\vsize=1.05\vsize
\voffset=-15true mm
\tolerance=10000

\newcount\footnum
\footnum=0
\def\fta{\relax}
\let\footold=\footnote
\def\footnote#1#2{\global\advance\footnum by 1
\footold{#1${}^{\the\footnum}$}{#2}}

\centerline {\bf RESONANCE TYPE INSTABILITIES}
\centerline {\bf IN THE GASEOUS DISKS}
\centerline {\bf OF THE FLAT GALAXIES}
\centerline {\bf III. The gyroscopical resonance type instability}
\medskip
\centerline {C.M.Bezborodov, J.V.Mustsevaya and V.V.Mustsevoy}
\bigskip

{\ff
The stability of thin homogeneous gaseous disk with step
discontinuity of velocity and density profile has been investigated
analytically in the limitary case of large compressibility and
numerically at spatially separated velocity and density kinks.
The expressions for reflection and transmission coefficients
at the axisymmetrical vortex sheet are derived. The possibility
of gyroscopical resonance type instability development in addition
to Kelvin--Helmholtz and centrifugal instabilities is shown
and its physical mechanism is clarified. It has been discussed
the possibility of application of obtained results to the stability
analysis of the gaseous disks of the real flat galaxies.
}

\medskip % ******************************************************
{ \it Introduction.}
\medskip % ******************************************************

The applied problems of astrophysics about stability of supersonic
currents (i.e. disks of accreting matter, galactic disks, jets
from young stars etc.) from the academic point of view are interesting
first of all by allowing to allocate many specific effects
and mechanisms of amplification of disturbances in the pure state.
This circumstances are related to absence of such complicating factors
as heat exchange with firm borders, non-absolute rigidity of these
borders, and also extremely small viscosity. Therefore it is
possible to combine successfully relative simplicity of models
with practical application of results obtained at their analys22-08-98 09:39pm22-08-98 09:39pmis.
In the present part of work with reference to gaseous disks of
flat (spiral) galaxies amplification of unstable disturbances
in a waveguide layer between jumps of rotation velocity and
density caused by over-reflection is discussed.

The effect of approaching infinity of reflection and transmission
coefficients of a flat monochromatic sound wave falling at certain
angles on a flat surface of supersonic vortex sheet (so called
the effect of over-reflection) was discovered by Miles (1957) and
Ribner (1957). Subsequently the similar effect was investigated
for magnetohydrodynamic jump (Acheson 1967). It was also shown that
the presence of jump is not obligatory for over-reflection and that
it also takes place on a critical layer in a flow with smooth velocity
distribution (Kolyhalov 1985) (that is on a layer where flow speed
and wave phase speed along flow speed coincide). The mechanism of
over-reflection is caused by that the energy flux in a transmitted
wave is directed to jump (Miles 1957, Ribner 1957). Thus the
transmitted wave outgoing from jump (or critical layer) picks up
kinetic energy from the basic current and transfers it to
a reflected wave through jump (see e.g. Landau \& Lifshits 1988).

Although Betchov \& Criminale (1967) marked that if to arrange
a reflecting surface (acoustic screen) in parallel to a critical
layer, an acoustical resonance type instability may develop.
However, their speculations did not go further than statement
of such opportunity. The first rather serious researches of
acoustic resonance type instability are probably works of Ferrari
et al. 1982, Payne \& Cohn 1985, Hardee \& Norman 1988 (see also
Norman et al. 1982), in which stability of astrophysical jets in
models of a flat layer and cylindrical jet with discontinuous
borders was investigated. The large number of works on the
specified theme have appeared subsequently but all of them
concerned to plainparallel gas flows.

The unstable modes caused by over-reflection from a vicinity of
corotation in gaseous disks with monotonous dependence of rotation
velocity on radius were found for the first time by Papaloisou and
Pringle (1985, 1987). Then such modes (and also other
resonant modes, for which amplification a continuity of rotation
velocity is essential) were investigated by Glatzel (1987),
Savonije \& Heemskerk (1990), Papaloizou \& Savonije (1991) etc.
Nevertheless as for disks with continuous radial parameters
distribution the linearized equations of gas dynamics can be
investigated only numerically, there was the number of unsolved
questions connected to dependence of characteristic time of
development of unstable disturbances and angular phase rotation
velocity of a pattern created by them upon unperturbed disk
parameters. Besides, in quoted works the consideration was
performed with reference to accretion disks.

In the present part of the article the effect of over-reflection and
instability caused by it in the elementary discontinuous model of a thin
gaseous disk is considered. Its applicability is caused by observable
peculiarities of density and rotation velocity distributions in disks
of flat galaxies.

The basis for consideration is such fact that the wave can be many
times reflected from rather sharp density jump and from corotation
radius, and from the latter with strengthening (Savonije \& Heemskerk
1990). It is clear that even when ``centres'' of smoothed jumps of
velocity and density distributions coincide, specified points of
turn generally speaking are separated on radius. Thus in a disk
the waveguide layer presents, in which wave collects energy
in due course.

Is the most simple model of such waveguide in a gaseous disk is model
of two jumps, namely vortex sheet\footnote {\fta}{\ff In discontinous
model corotation radius with necessity coincides with vortex sheet,
because phase speed of a unstable wave should lay between the minimum
and maximum speeds of the basic current, that it is possible to prove
with the help of a theorem (Kolyhalov 1985).}
and internal rather it contact jump. The choice of discontinuous
model for the analysis requires some explanations, as to the present
there is the number of works where stability of gaseous disks with
continuous velocity and density profiles (see, for example,
Papaloizou \& Pringle 1985, 1987, Glatzel 1987, Savonije \& Heemskerk
1990, Papaloizou \& Savonije 1991, Torgashin 1986, Morozov 1989,
Morozov et al. 1992) was analyzed.

However in a disk with continuous parameters distributions the
instability as a rule is developed simultaneously by several
mechanisms (for example, over-reflection and over-transmission
of waves in a corotation vicinity (Savonije \& Heemskerk 1990),
resonant radiation of waves from corotation radius (ibidem),
centrifugal (Torgashin 1986, Morozov 1989, Morozov et al. 1992),
centrifugal-resonant (Morozov et al. 1992), Kelvin--Helmholtz
instability (KHI) mechanism (Torgashin 1986) etc.).
So it is almost impossible to tell them from each other and
to reveal their relative influence. In the contrary
the discontinuous model will allow us in section 3 to show
that instability developing in it is of a really new type.
Though the physical mechanism of this instability has similar
features with the mechanism of acoustic resonance type
instability in plainparallel current (Payne \& Cohn 1985),
there is conceptual difference consisting that a resonance
occurs rather at sound frequency than at gyroscopic one.
Accordingly frequency areas of unstable modes localization are
determined not by ``sound'' frequency $|{\bf k}| c_s$ but
gyroscopic one $\Omega_{in; ex}(m \pm 2) $, where $m$ is
azimuthal mode number (number of an arms of spiral pattern).
Besides such model will allow to allocate in the pure state
unstable superficial modes: centrifugal connected with velocity
jump and mode connected to density drop, and to investigate
dependence of their properties on parameters of a disk.
Thus an independent problem is to determine the expressions
for wave falling on axisymmetic vortex sheet reflection and
transmission coefficients (section 2), as though they have large
importance for our consideration but were not received former.

Finally, it seems useful to discuss in section 4 the probable
astrophysical applications of obtained results.

In the present part of the work we shall use the references to
the formulae from Parts I and II, adding in this case a symbol
``1.'' or ``2.'' accordingly before number of the formula.

\medskip% **************************************************************
{\it 1. Model and dispersion equation.}
\medskip % **************************************************************

The basic characteristic features of model used in the present part
of our work are the same as in model described in a Part II. We will
therefore stop only at differences.

The model contains not one but two jumps: contact jump at radius
$R_\rho$ and vortex sheet at $R_\Omega$. Let's assume for
definiteness $R_\rho < R_\Omega$ as it is seems such situation
is more characteristic for real objects (at least it is valid
for the Galaxy). The radial dependences of unperturbed parameters
are set by relations:
$$\eqalign{
\rho_0&(r, z) = \rho_{in}(z) +
	(\rho_{ex}(z) - \rho_{in}(z)) \,\theta (r-R_\rho), \cr
c_s&(r, z) = {c_s}_{in} (z) \left[1 +
   \left( \sqrt {\rho_{in} (z) \over\rho_{ex} (z)} - 1 \right)
   \theta(r-R_\rho) \right], \cr
\Omega&(r, z) = \Omega_{in}(z) +
	(\Omega_{ex}(z) - \Omega_{in}(z)) \,\theta(r-R_\Omega), \cr
P_0&(r, z) = c_s^2 (r, z) \rho_0(r, z) /\gamma = const(z). \cr} \eqno (1)
$$

As the magnitudes of disturbances should be limited in zero and
at infinity, the solution of the appropriate modified Bessel
equation in the given model takes a form:
$$ p (r) = \left\{
\matrix {
A I_m (k_{in} r), & \quad r < R_\rho, \cr
B I_m (k_{me} r) + C K_m (k_{me} r), & R_\rho < r < R_\Omega, \cr
D K_m (k_{ex} r), & \quad r> R_\Omega. \cr
} \right.							\eqno (2)
$$
Here $k_{in}$ and $k_{ex}$ are defined by (2.6) and in area
between jumps
$$ k^2_{me} = {4\Omega^2_{in} - (\omega
	- m\Omega_{in})^2 \over {c^2_s}_{ex}}. 			\eqno (3)
$$

It is obvious the conditions at jumps do not differ from (2.8), (2.9):
$$\eqalign{
p (R_\rho + 0) - p (R_\rho-0) =& 0, \cr
p (R_\Omega + 0) - p (R_\Omega-0) =& \rho_{ex}\,\xi (R_\Omega) \,R_\Omega
	(\Omega^2_{in} - \Omega^2_{ex} ), \cr
\xi (R_\rho + 0) - \xi (R_\rho-0) =&
	\xi (R_\Omega + 0) - \xi (R_\Omega - 0) = 0. \cr
}								\eqno (4)
$$
Substituting the solution as (2) in conditions (4) we come to the
system of linear algebraic equations on coefficients $A, B, C, D$.
A condition of simultaneity of this system will be equality to zero
of its determinant. Excepting neutral gyroscopic modes of
fluctuations $\omega = (m\pm 2) \Omega_{in}$, $\omega = (m\pm 2)\Omega_{ex}$
we finally write:
$$\left|
\matrix {
1
	& 1
	& 1
	& 0 \cr
\beta^{(\rho)}_{in}
	& Q\alpha^{(\rho)}_{me}
	& Q\beta^{(\rho)}_{me}
	& 0 \cr
0
	& \delta_K
	& \delta_I
	& 4q^2- (x-mq)^2 - \alpha^{(\Omega)}_{ex} (1-q^2) \cr
0
	& \delta_K \alpha^{(\Omega)}_{me}
	& \delta_I \beta^{(\Omega)}_{me}
	& \alpha^{(\Omega)}_{ex} [4 - (x-m)^2)] \cr
} \right| = 0. 							\eqno (5)
$$
The equality (5) represents the dispersion equation for disturbances
in considered model. Designations of Parts I and II in combinations
$\alpha^{(j)}_i$ and $\beta^{(j)}_i$ setting by (1.8) and (1.9)
are here kept; top index specifies that they are written on radius
of velocity or density jump, the bottom index takes values
``$in$'', ``$ex$'' or ``$me$'' (with the account
$\Omega_{me}\equiv\Omega_{in}$). Besides following designations
are introduced:
$$
  \delta_I = {I_m (k_{me}R_\Omega) \over I_m (k_{me}R_\rho)};\quad
  \delta_K = {K_m (k_{me}R_\Omega) \over K_m (k_{me}R_\rho)}.	\eqno (6)
$$

Before solving directly the dispersion equation (5) in the following
section we consider a problem on reflection of a neutral wave
from axisymmetic vortex sheet to have an opportunity to apply
its results at interpretation of the instability mechanism.

\bigskip % **************************************************************
{ \it 2. Reflection and transmission coefficients.}
\medskip % **************************************************************

From theoretical works (Miles 1957, Ribner 1957) it is known that
at incidence on plainparallel vortex sheet the neutral wave can
reflect and refract with amplification on amplitude (over-reflection
effect). At presence of the second reflecting surface (for example,
density jump or kink) spatial resonator will be formed, in which
the wave can collect energy in due course. Therefore the over-reflection
is capable to result in development of instabilities of
a waveguide-resonant type, in particular gyroscopic resonance type
instability considered in the given section.

For plainparallel shift currents the problem of over-reflection
up to now is investigated well even for systems with magnetic
fields (Acheson 1967). However, coefficients of reflection and
transmission for a neutral wave falling on axisymmetic jump
were not obtained till now.

Note at once that in cylindrical geometry there is the allocated
direction on centre. It leads to distinction of coefficients for
waves falling on axisymmetic jump from within and from outside.
Nevertheless, in a limit of large jumps radii or azimuthal
short-wave disturbances the expressions for coefficients received
in our model should lead to similar expressions from
a plainparallel problem.

Following Miles (1957) and Ribner (1957), we shall consider incidence
of a neutral monochromatic wave on vortex sheet being not interested
in this section by a question of its stability (on formal correctness
of such approach see Landau \& Lifshits 1988).

Further in this section the index ``$i$'' concerns to a incident wave,
``$r$'' --- to reflected, index ``$t$'' --- to transmitted ones.
The magnitudes in internal area from jump are designated by index $1$,
outside --- $2$.

\medskip
{\it a) Incidence from within on density and velocity jump}

By applying results of a Part II and condition of limitary
absorption (i.e. the disturbances should fade as they propagate)
at once write out peak functions of pressure of all three waves:
$$
   p_i = A K_m (k_1 r) ;\quad p_r = B I_m (k_1 r) ;\quad
   p_t = C K_m (k_2 r). 					\eqno (7)
$$
Thus displacement magnitudes on dividing border are
$$\eqalignno{
\xi_1 = \xi_i + \xi_r =& {A K_m (k_1 R) \alpha_1 +
	B I_m (k_1 R) \beta_1 \over \rho_1 {c_s^2}_1 k_1 R}, & (8) \cr
\xi_2 = \xi_t =&
	{C K_m (k_2 R) \alpha_2 \over \rho_2 {c_s^2}_2 k_2 R}. & (9) \cr}
$$
The parameters $\alpha_i,\ \beta_i$ are governed by the formulae
(1.8), (1.9).

Now introduce complex reflection and transmission coefficients
with respect to pressure by a usual way:
$$
\hbox{\sy R}^{(-)} \equiv
	{p_r\over p_i} = {B I_m(k_1R) \over A K_m(k_1R)};\quad
\hbox{\sy T}^{\,(-)} \equiv
	{p_t\over p_i} = {C K_m(k_2R) \over A K_m(k_1R)}. 	\eqno (10)
$$

Taking into account general conditions (2.8), (2.9) come to the equations on
$\hbox{\sy R}^{(-)},\ \hbox{\sy T}^{\,(-)}$:

$$ {k^2_1R^2\over k^2_2R^2} {\alpha_2\over\alpha_1}
	\hbox{\sy T}^{\,(-)} - {\beta_1\over\alpha_1}
	\hbox{\sy R}^{(-)} = 1, 				\eqno (11)
$$
$$ {\mu^2\hbox{\sy T}^{\,(-)}\over Q} \left\{1 -
	{M^2 (1 + Q) \over2\mu^2} {\alpha_2\over k_2^2R^2}
	(1-q^2) \right\} = 1+\hbox{\sy R}^{(-)}. 		\eqno (12)
$$

Solving this system let us write coefficients in an obvious kind:

$$\eqalignno{
   \hbox{\sy R}^{(-)} =& {Qk_1^2R^2\alpha_2 - \mu^2k_2^2R^2\alpha_1 +
{ 1\over2} (1 + Q) (1-q^2) M^2\alpha_1\alpha_2 \over
   \mu^2 k_2^2 R^2 \beta_1 - Q k_1^2 R^2\alpha_2 -
{ 1\over2} (1 + Q) (1-q^2) M^2\beta_1\alpha_2}, &(13) \cr
   \hbox{\sy T}^{\,(-)} =& {Q k_2^2 R^2 (\beta_1 - \alpha_1) \over
   \mu^2 k_2^2 R^2\beta_1 - Qk_1^2 R^2\alpha_2 -
{ 1\over2} (1 + Q) (1-q^2) M^2\beta_1\alpha_2}. &(14) \cr
} $$

\medskip

{ \it b) Incidence from outside on density and velocity jump}

In a case when the neutral wave falls on vortex sheet from
the large radii the solution of modified Bessel equation
takes a form:
$$
   p_i = A I_m (k_2 r); \quad p_r = B K_m (k_2 r) ;\quad
   p_t = C I_m (k_1 r). 					\eqno (15)
$$
With the help (15) write down displacement on jump:
$$\eqalignno{
\xi_1 = \xi_t =& {C I_m (k_1R_\rho) \beta_1 \over
   \rho_1 {c_s^2}_1 k_1^2 R}, 					&(16) \cr
\xi_2 = \xi_i + \xi_r =& {A I_m (k_2 R) \beta_2 +
B K_m (k_2 R) \alpha_2 \over\rho_2 {c_s^2}_2 k_2^2 R} 		&(17) \cr
} $$
Applying again the condition on jump and making simple transformations
we obtain coefficients $\hbox{\sy R}^{(+)}$ and $\hbox{\sy T}^{\,(+)}$:
$$\eqalignno {
\hbox{\sy R}^{(+)} =& {\mu^2 k_2^2 R^2 \beta_1 - Qk_1^2 R^2 \beta_2 -
{ 1\over2} (1 + Q) (1-q^2) M^2 \beta_1 \beta_2 \over
   Qk_1^2 R^2\alpha_1 - \mu^2 k_2^2 R^2\beta_1 +
{ 1\over2} (1 + Q) (1-q^2) M^2\beta_1\alpha_1}, 		&(18) \cr
\hbox{\sy T}^{\,(+)} =& {\mu^2 k_1^2 R^2 (\alpha_1 - \beta_2) \over
	Qk_1^2 R^2\alpha_1 - \mu^2 k_2^2 R^2\beta_1 +
{ 1\over2} (1 + Q) (1-q^2) M^2\beta_1\alpha_1}. 		&(19) \cr
} $$

Note that classical connection between {\sy R} and {\sy T}
({\sy T}$= 1\,+${\sy R}) known for a plainparallel problem
(Miles 1957, Ribner 1957, Landau \& Lifshits 1988) in our case
is valid on density jump only as just on it the boundary
conditions coincide identically with those of a flat problem,
namely both $\xi$ and $p$ are continuous.

\bigskip
It is interesting to note taking the determinant (5) it is
possible to write dispersion equation through the reflection
coefficients production:
$$ \hbox{\sy R}_\rho^{(+)}\hbox{\sy R}_\Omega^{(-)}
	= {I_m (k_{me}R_\Omega) \over I_m (k_{me}R_\rho)}
        {K_m (k_{me}R_\rho) \over K_m (k_{me}R_\Omega)}. 	\eqno (20)
$$
Here the index ``$\rho$'' marks reflection coefficient on density jump
($R =R_\rho,\ q=1$) and index ``$\Omega$'' --- on velocity jump
($R = R_\Omega,\ Q=\mu=1$). The latter equation allows to demonstrate
evidently that reflection coefficients are associated with the
instability mechanism that we consider in the following section.
Really, in the limit of short radial wavelength
$(|k_{me}R_\rho|\gg m)$ from (20) we find:
$$ \hbox{\sy R}_\rho^{(+)}\hbox{\sy R}_\Omega^{(-)}
	\simeq {\rm exp} \left\{2k_{me} (R_\Omega-R_\rho) \right\}
	\simeq {\rm exp} \left\{-2iD {M\over\mu} (x-m) \right\}. \eqno (21)
$$
Dimensionless width of a backlash between jumps is here intoduced:
$$ D = {R_\Omega - R_\rho \over R_\Omega}. $$
Taking a square of the (21) module we get:
$$ \left|\hbox{\sy R}_\rho^{(+)}\hbox{\sy R}_\Omega^{(-)}\right|^2 \simeq
{\rm exp} \left\{4D {M\over\mu} Im\, x \right\} =
{\rm exp} \left\{4 {(R_\Omega-R_\rho) \over{c_s}_{ex}} Im\,\omega \right\}.
								\eqno (22)
$$
Thus for unstable ($Im\,\omega > 0$) disturbances
$\left| \hbox{\sy R}_\rho^{(+)}\hbox{\sy R}_\Omega^{(-)}\right| >1$
is always valid. From (22) follows
$$ Im\,\omega \simeq {1\over2\tau_c} {\rm ln} \left|
\hbox{\sy R}_\rho^{(+)}\hbox{\sy R}_\Omega^{(-)}\right|, 	\eqno (23)
$$
where $\tau_c$ is characteristic run-time of a sound wave between
jumps. Note that the result (23) is typical for models admitting
development of acoustic resonance type instabilities (Payne \& Cohn 1985).

\medskip% **************************************************************
{ \it 3. The stability analysis.}
\medskip% **************************************************************

Before receiveing the approximated solutions of the equation (5)
it is necessary to make one remark. For existance of over-reflection
and unstable reflective harmonics caused by it a nonzero wave
energy flux in a radial direction is necessary at least in two areas
(at $r>R_\Omega$ and at $R_\rho < r < R_\Omega$) the same as and
in a flat case (see Morozov et al. 1991). It is possible if
the eigenfunctions $p$ and $\xi$ in these areas are oscillating
on radial coordinate (as for value of group velocity a value of
radial component of wave vector $k_r$ is crucial). The latter
may occur only when arguments of modified Bessel functions
are imaginary (the preceeding concernes to a limit
$Im\,\omega \rightarrow 0$, as for such disturbances only
an energy flux is defined). Writing $k^2 = -k_r^2$ in each area
we come to restriction on a real part of frequency, namely,
the following inequalities should {\it not} occur:
$$ (m-2) \Omega_{in} \le Re\,\omega \le (m + 2) \Omega_{in}, 	\eqno (24)
$$
$$ (m-2) \Omega_{ex} \le Re\,\omega \le (m + 2) \Omega_{ex}. 	\eqno (25)
$$
As for increasing disturbances
$$
	m\Omega_{ex} \le Re\,\omega \le m\Omega_{in} 		\eqno (26)
$$
is valid (at $\Omega_{ex} < \Omega_{in}$), the reflective
harmonics can take place only in following frequency range:
$$
   (m + 2) \Omega_{ex} < Re\,\omega < (m-2) \Omega_{in},	\eqno (27)
$$
that it is possible at $m > 2$. Note that here is the strict
inequality, as neutral gyroscopic modes have already been
excluded from consideration.

Will jump-orthogonal component of a wave vector be real or
imaginary (and accordingly presence or absence of energy flux
in this direction) is determined in a flat case by a relation
of squares of Doppler $\omega-{\bf k}_\parallel{\bf V} $
and sound $k_\parallel c_s$ frequencies, and in considered
case --- by Doppler $\omega-m\Omega_i$ and gyroscopic
$2\Omega_i$ ones. And in the latter case changing $c_s$
simply scales $k$ ($|k| \to\infty$ at $c_s \to 0$ and
$|k| \to 0$ at $c_s\to\infty$), not breaking proportions between
$Im\,k$ and $Re\, k$. Accordingly the resonance will take place
on harmonics of gyroscopic frequency $(m-2) \Omega_{in}$ in
a layer of gas between velocity and density jumps. This allows
on the one hand to limit ourselves in searching of reflective
harmonics by permitted area of frequencies (27) (this area
is limited by continuous lines on Fig.~3.1) and on the other
hand to set the value $(m-2)\Omega_{in}$ as initial approximation
of frequency of the basic unstable mode in a waveguide
layer between jumps.

\medskip
Let us obtain the solution of the dispersion equation (5) in
a limit of strong compressibility of media in all three areas
divided by jumps ($M=R_\Omega \Omega_{in} /{c_s}_{in}\gg1,\ M/\mu\gg1$).
Remember, magnitude of arguments of modified Bessel functions
is determined in the basic by Mach number. It allows to put
$|k_i R_\rho| \gg m$; in this case the dispersion equation
is simple to bring to a kind:
$$ {Q\ {\rm th}\, a + \mu\over \mu\ {\rm th}\, a + Q}
   \simeq {x-m \over - (x-mq) + i {M\over\mu} (1-q^2)}, 	\eqno (28)
$$
where $a = k_{me}R_\Omega\,D = {MD\over\mu} \sqrt {4- (x-m)^2} $.
For the case of uniform on density at $z = const$ disk
($Q=\mu=1$) from (28) result (2.12) describing centrifugal instability
mode directly follows. In general case ($Q\ne1$) if only $D\to0$ has not
place with accuracy to the terms of order
$| e^{-2a} (Q-\mu) / (Q + \mu) | \ll1$ we find:
$$
x \simeq {1\over2} m (1 + q) + i {M\over2\mu} (1-q^2). 		\eqno (29)
$$
Expression (29) means that if the jumps are not too close
the centrifugal mode of velocity jump in the main order
is not ``sensitive'' to the density jump.

For the further calculations transform (28) as follows:
$$
{\rm th}\, a \simeq - {Q (x-m) + \mu (x-mq) - iM (1-q^2) \over
   \mu (x-m) + Q (x-mq) - i {MQ\over\mu} (1-q^2)}. 		\eqno (30)
$$

In a case of very close incoincident jumps ($D\ll 1$) the left-hand
part of (30) can be presented as: ${\rm th}\,a \simeq - iMD (x-m)/\mu$.
Substituting this result in (30) we get the square equation on $x$,
but despite its simplicity the solution appears extremely inconvenient
for the analysis because of bulkness. Therefore let us limit by a case
$D\to0$ at which we find:
$$ x \simeq {1\over\mu + Q} \left\{m (Q + q\mu) +
   iM (1-q^2) \right\}. 					\eqno (31)
$$
It is easy to note, that (31) does not coincide with the asymptotic
solution (2.11) for concurrent jumps. Thus the limiting transition
to a case $D\equiv0$ is absent. It is a direct consequence that
we work with model of ``jump'' with finite thickness
($\xi\ll\Lambda_\Omega,\,\Lambda_\rho$, where
$\Lambda_\Omega,\,\Lambda_\rho$ are characteristic radial scales
of jumps of angular velocity and density), in which as it was
specified by Fridman \& Khoruzhii (1993) the large role
is played with a specific kind of equilibrum parameters inside
``jump'' profile. It is obvious, the result (31) could be obtained
also in model of concurrent jumps if to set ``asymmetrical''
concerning middle of ``jump'' $\rho(r)$ and $\Omega(r)$
distributions that would change boundary conditions (Fridman \&
Khoruzhii 1993). The physical sense of (31) becomes perfectly
transparent if to write growth rate (31) in the dimensional form:
$$
Im\,\omega = {\rho_{ex}R_\Omega (\Omega^2_{in} -\Omega^2_{ex})
\over{c_s}_{in}\rho_{in} + {c_s}_{ex}\rho_{ex}}. 		\eqno (32)
$$
So the converce characteristic time of instability development
is determined by the ratio of centrifugal force density drop
on angular velocity jump to the sum of media wave resistances
on both sides from a layer containing both jumps.

\medskip

To find out reflective harmonics we transform the left-hand
part of (30) by presenting dimensionless frequency as
$x = x_0 + \delta x$ and $ | \delta x |\ll x_0$:
$$
{\rm th}\left( {MD\over\mu} \sqrt{4 - (x-m)^2} \right)
	\simeq i\,{\rm tg}\left[{MD\over\mu} \sqrt{(x_0-m)^2-4}
	+ C\,\delta x \right],					\eqno (33)
$$
where
$$ C = {MD\over\mu} {(x_0-m) \over\sqrt {(x_0-m)^2-4}}. $$

Transforming (33) further, we assume
${MD\over\mu} \sqrt{(x_0-m)^2-4} = \pi n$.

Then (33) degenerates to $i\,{\rm tg}(\pi n + C\,\delta x) =
i\,{\rm tg}(C\,\delta x) $. Making the additional assumption
$|C\,\delta x| \ll 1$ we get ${\rm tg}(C\,\delta x) \simeq C\,\delta x$,
with respect of told let us write approximated expression
for frequency of reflective harmonics (number of harmonic $n=1,2,3,...$
determines number of zeros of eigenfunctions between jumps):
$$ x_0 = m - \sqrt {4 + {\left(\mu\pi n\over MD\right)}^2};	\eqno (34)
$$
$$ \delta x \simeq {m (Q + \mu q) - x_0 (Q + \mu)
	+ iM (1-q^2) \over ( Q + \mu) + {CMQ\over\mu} (1-q^2)
	+ iC\left[x_0 (Q + \mu) - m (\mu + qQ) \right]}.	\eqno (35)
$$
Remind, by virtue of told in the section beginning, $x_0$ lays in limits:
$$
	(m + 2) q < x_0 < (m-2). 				\eqno (36)
$$
At last let us find approximated expression for basic (with $n=0$)
unstable waveguide mode. Note, by simple substitution $n=0$ in
(34), (35) the frequency of this mode cannot be described as then
$C$ turns to infinity.

We believe $x \sim m-2$. Then in approximation of $|k_{in}R_\rho|,
|k_{me}R_\Omega| \ll 1,\ |k_{ex}R_\Omega| \gg 1$ substituting
frequency as $x = m - 2 + \delta x$ in the equation (5) we determine:
$$
\delta x \simeq {4\left[1-q^2 + {i\mu\over M} (m-2-mq) \right]
	\over {8\over m} \left[1 - \Xi\right] -
	{4i\mu\over M} + \left[(1 + Q) + \Xi \right]
	\left(1-q^2 + {i\mu\over M} (m-2-mq) \right)},		\eqno (37)
$$
where $\Xi \equiv (1-Q) (1-D)^{2m}$.

Remarkable peculiarity of expressions (34), (35), (37) is the fact
that they describe unstable roots both at $\Omega_{in} > \Omega_{ex}$
and at  $\Omega_{ex} > \Omega_{in}$ and in an essentially supersonic
case ($R_\Omega |\Omega_{in}-\Omega_{ex}| /{c_s}_{ex} \gg1$),
and are similar to analogous asymptotics of reflective harmonics
and basic mode developing between density and velocity jumps in
a plainparallel flow (see Morozov et al. 1991). According to told
in the beginning of section instability described by expressions
(34), (35), (37) we shall name the gyroscopic resonance type
instability (GRTI). The absence of centrifugal suppession of
instability at $\Omega_{ex} > \Omega_{in}$ is caused by overreflective
character of its development mechanism that is by specifically
wave effect not connected directly to mass forces density
distribution in a disk.

The case of small compressibility ($M\ll1$) cannot be investigated
analytically because of bulkness of the approximated algebraic
equation obtained in this limit.

\medskip
The results of the numerical solution of (5) (see Fig.~3.2--3.7)
as a whole confirm conclusions made on the basis of asymptotic
analytical research and at essential compressibility
$(R_\Omega(\Omega_{in}-\Omega_{ex})/{c_s}_{ex}\gg1)$
and small relative distance between jumps $D\ll 1$
these results show presence besides mode of centrifugal
instability a discrete set of poorly unstable GRTI modes.
The phase angular velocity of rotation of a pattern of these modes
$\Omega_p=Re\,\omega/m$ extremely weakly depends on $Q$ and $q$
(Fig.~3.2, 3.4) and essentially changes with $D$ and $M$ change
(Fig.~3.5, 3.6), remaining thus in permitted frequency area (27)
(see also Fig.~3.1). Exception makes basic (not having zeros of
eigenfunctions between jumps) GRTI mode for which in all
parameters ranges $Re\,\omega$ unsignificantly exceeds
$(m-2)\Omega_{in}$. The number of reflective ($n\ge 1$) harmonics
is increased with growth of Mach number, azimuthal mode
number $m$ and distance between jumps reduction.

As it is seen from Fig.~3.8 the asymptotic solution of (34), (35)
gives only qualitative conformity to numerical results.
It is connected that at approximated taking square roots in
$k_iR_\Omega$ we have neglected a square of gyroscopic frequency
in comparison with a squared Doppler one, whereas the effect of
over-reflection resulting in GRTI occurrence is caused by
a resonance just between these frequencies. Nevertheless,
(34), (35) begin to work well at large numbers $m$ when
the calculation of modified Bessel functions on the
computer is complicated.

Let us specify at last, that as well as in a Part II the considered
disturbances with $\tilde v_z \equiv 0$ are allocated by a natural
way by the instability mechanism itself. Really, in this sense
all told in Part II is valid for centrifugal instability mode,
and resonant harmonics with a wave vector componenet $k_z$
comparable with $k_r$ and $k_\varphi$ during the time of $h/c_s$
order (here $h$ is the characteristic disk half-thickness) leave
a waveguide layer. As this time is near to inverted growth rate
of resonant modes, such disturbances will have no time to amplify
up to appreciable amplitudes. If to take into account
an opportunity of returning of these disturbances to a waveguide
layer because of refraction, their growth rate anyway will be
less than at disturbances not having $k_z$.

\medskip% **************************************************************
{ \it 4. Conclusions.}
\medskip% **************************************************************

Let us generalize numerical and analytical results by formulating
the conclusions about stability of a uniform on pressure at
$z = const$ gas disk containing rotation angular velocity jump
and internal rather it non-homentopic density jump (in models
of ``jumps'' of finite thickness, that is
$\xi\ll\Lambda_\Omega,\,\Lambda_\rho$,
where $\Lambda_\Omega,\,\Lambda_\rho$ are characteristic radial
scales of angular velocity and density jumps).

1. In a limit of weak compressibility in considered system only
one unstable mode resulting Kelvin--Helmholtz instability
development on vortex sheet can develop.

2. With growth of compressibility the development mechanism
of the vortex sheet superficial mode essentially changes
from Bernoulli effect resulting in Kelvin--Helmholtz instabilities
up to centrifugal one in a supersonic case.

3. The presence of density jump in the system appreciably
impacts to centrifugal instability development only when
the jumps are very close, i.e. $D\ll1$.

4. At essentially supersonic velocity difference on jump and
$m > 2$ a discrete set of weakly unstable modes appears.
These modes differ by number of zeros of eigenfunctions between
jumps. The growth rate of these modes appreciably differs
from zero only when $D\ll1$ and their amount grows with growth of
compressibility on jump, azimuthal mode number, with distance
between jumps reduction and practically does not depend on
other parameters. These modes are caused by over-reflection
on velocity jump and collect energy in a waveguide layer
between jumps exponentially in time, picking it up from kinetic
energy of the main movement (rotation) of outer gas layer.

5. The effect of over-reflection is caused by a resonance between
Doppler and gyroscopic frequencies\footnote {\fta}{\ff The last in
considered model with discontinuous velocity profile represents
a degenerate case of epicyclic frequency.} and the waveguide
unstable modes described above are displayed near to the basic
gyroscopic frequency and its higher harmonics. So the given
instability is named as gyroscopic resonance type instability (GRTI).

6. As against centrifugal instability the GRTI develops and at
``negative'' velocity drop on jump (when the periphery of a disk
rotates faster than the central region).

\bigskip

Finishing the article, we make some final remarks on probable
applied meaning of obtained results.

It is necessary to note, GRTI by virtue of small scale and small
growth rates is hardly capable to compete to centrifugal and
Kelvin--Helmholtz instabilities as the generator of large-scale
spiral structure, its role rather can be reduced unless to
creation of multi-arm secondary (imposed on basic) pattern
(see Morozov et al. 1992). At the same time it is known,
the similar resonant Papaloisou--Pringle instabilities and
acoustic resonance type ones are much more weakly subject
to stabilizing influence of velocity jump smoothing and besides
are multimode (Savonije \& Heemskerk 1990, Papaloizou \& Savonije
1991, Mustsevoy \& Hoperskov 1991). Add that the resonant modes
by virtue of its development slowness do not result to
rough rearrangement of system parameters distributions and
to system exiting on stability margin. Therefore the simultaneous
development of large number of various harmonics of different
azimuthal GRTI modes probably is capable to lead to effective
removal of an angular moment from a vicinity of a waveguide
layer on disk periphery and to smooth durable change of radial
profiles of density and rotation velocity.

Make at last one more remark on possible GRTI and other
hydrodynamic resonant instabilities influence on magnitude
of stars velocities dispersion in a galactic disk.

Besides gyroscopic resonance type instability considered in
the previous sections the development of other resonant
instabilities (e.g. Papaloisou-Pringle instability (Papaloizou 1985,
Savonije \& Heemskerk 1990), resonant-centrifugal one (Morozov et al.
1992)) is possible in a differentially rotating gaseous disk.
All of them are multimode and are characterized by discrete sets
of radial, azimuthal and transverse to a disk wave numbers and
frequency eigenvalues. Thus in result of development of resonant
type instabilities in a gaseous galactic disk there is the complex
system of disturbances of velocity and thermodynamic parameters
with hierarchy of spatial and temporary scales. It is rather
essential that the energy of these disturbances exceeds sound
energy or tends to it from above. The basis for the last statement
give laboratory (Norman et al. 1982) and nonlinear numerical
(Norman \& Hardee 1988) modelling of supersonic jets.
In these experiments it is shown that the growth of disturbances
caused by resonant instabilities stops at a nonlinear stage at the
formation of system of weak slanted shock waves (disturbance with
wavelength comparable to characteristic jet size) on a background
of advanced acoustic turbulence (the most short wavelength
disturbances). On the other hand it is known that in galactic disks
rich by gas ($\sim 10\%$), the gravitational stability of a disk
is determined by more ``cold'' gas componenet (Jog \& Solomon 1984).
Therefore in numerical experiments on modeling binary (gas + stars)
disk the gravitational instability of a gas subsystem develops
and results in formation of giant molecular complexes (GMC).
In turn relaxational processes (stars scattering on massive GMC)
result to fast ``warm up'' of a star disk. As a result stars
velocity dispersion becomes so high that the star disk appears
gravitationally stable even at a unstable initial condition
(described scenario was observed, for example, in experiments of
Shlosman \& Noguchi 1993). So in numerical experiments
the specified process could result in a conclusion of
impossibility of spiral structure or bar generation by
development of gravitational instability of a star disk.

As the margin of gravitational stability is determined by
balance of energy of an ordered movement under action of
gravitational forces and stochastic thermal movements and
as the energy of small-scale system of weak slanted shock waves
caused and supported by multimode resonant instabilities,
dissipates in thermal because of viscosity (though small),
it is necessary to make a conclusion that the really gas disk
in an extended vicinity of a waveguide layer is warmed up
and consequently appears to be more gravitationally stable.
Besides velocity and density fluctuations caused by resonant
instabilities should slow down GMC formation, partially to destroy
already formed complexes and accordingly make scattering of
stars on them less effective.

So the account of the really working resonant type instabilities
ought cause reduction of stars velocities dispersion in numerical
modeling ``gas + stars''. We have used a subjunctive inclination
as the resonant modes conceptually are short-wave and cannot be
observed in really feasible experiments as their
characteristic scales are less or comparable to the
characteristic size of a cell and they are inevitably absorbed
by numerical viscosity.

\medskip
{\it Acknowledgement.} One of us (VVM) is grateful to INTAS
for support of this work by grant project N 95-0988.

\bigskip % **************************************************************
\centerline {\ref References}
\medskip% **************************************************************

Acheson, D.J., 1967, J. Fluid Mech., 77, 433 \par
Betchov, R., \& Criminale, W.O., 1967, Stability of Parallel Flows
	(New York: Academic Press) \par
Ferrari, A., Massaglia, S., \& Trussoni, E., 1982, MNRAS, 198, 1065 \par
Fridman, A.M., \& Khoruzhii, O.V., 1993, Sov. Phys. Uspekhi, 163, 79 \par
Glatzel, W., 1987, MNRAS, 225, 227 \par
Hardee, P.E., \& Norman, M.L., 1988, Ap. J., 334, 70 \par
Jog, C.J., \& Solomon, P.M., 1984, Ap. J., 276, 114 \par
Kolyhalov, P.I., 1985, Sov. Phys. Dokl., 180, 95 (in Russian) \par
Landau L.D., Lifshits E.M., 1986, Gidrodinamika (Hydrodynamics),
	3rd ed. (Moscow: Nauka) (There exist an English edition, 1987). \par
Miles, J.W., 1957, J. Acoust. Soc. Amer., 29, 226 \par
Morozov, A.G., 1989, KFNT, 5, 75 (in Russian) \par
Morozov, A.G., Mustsevaya, J.V., \& Mustsevoy, V.V., 1991,
	Preprint of Volgograd State Univ., 2-91 \par
Morozov, A.G., Mustsevoy, V.V., \& Prosvirov, A.E., 1992,
	SvA. Lett., 18, 46 \par
Mustsevoy, V.V., \& Khoperskov, A.V., 1991, SvA. Lett., 17, 281 \par
Norman, M.L., Smarr, L., \& Vinkler, K.-H., 1982,
	in ``Numerical Astrophysics. Proceeedings of a symposium
	in honor of James R.~Wilson...'' (Boston: Jones and Bartlett
	Publishers, Inc.) \par
Norman, M.L., \& Hardee, P.E., 1988, Ap. J., 334, 80 \par
Papaloizou, J.C.B., \& Pringle, J.E., 1985, MNRAS, 213, 799 \par
Papaloizou, J.C.B., \& Pringle, J.E., 1987, MNRAS, 225, 267 \par
Papaloizou, J.C.B., \& Savonije, G.J., 1991, MNRAS, 248, 353 \par
Payne, D.G., \& Cohn, H., 1985, Ap. J., 334, 80 \par
Ribner, H.S., 1957, J. Acoust. Soc. Amer., 29, 435 \par
Savonije, G.J., \& Heemskerk, M.H.M., 1990, A\&A, 240, 191 \par
Shlosman, I., \& Noguchi, M., 1993, Ap. J., 414, 474 \par
Torgashin, Yu.M., 1986, PhD diss., Tartu (in Russian) \par

\vfil\eject

\bigskip
\epsfxsize=\hsize
\epsfbox{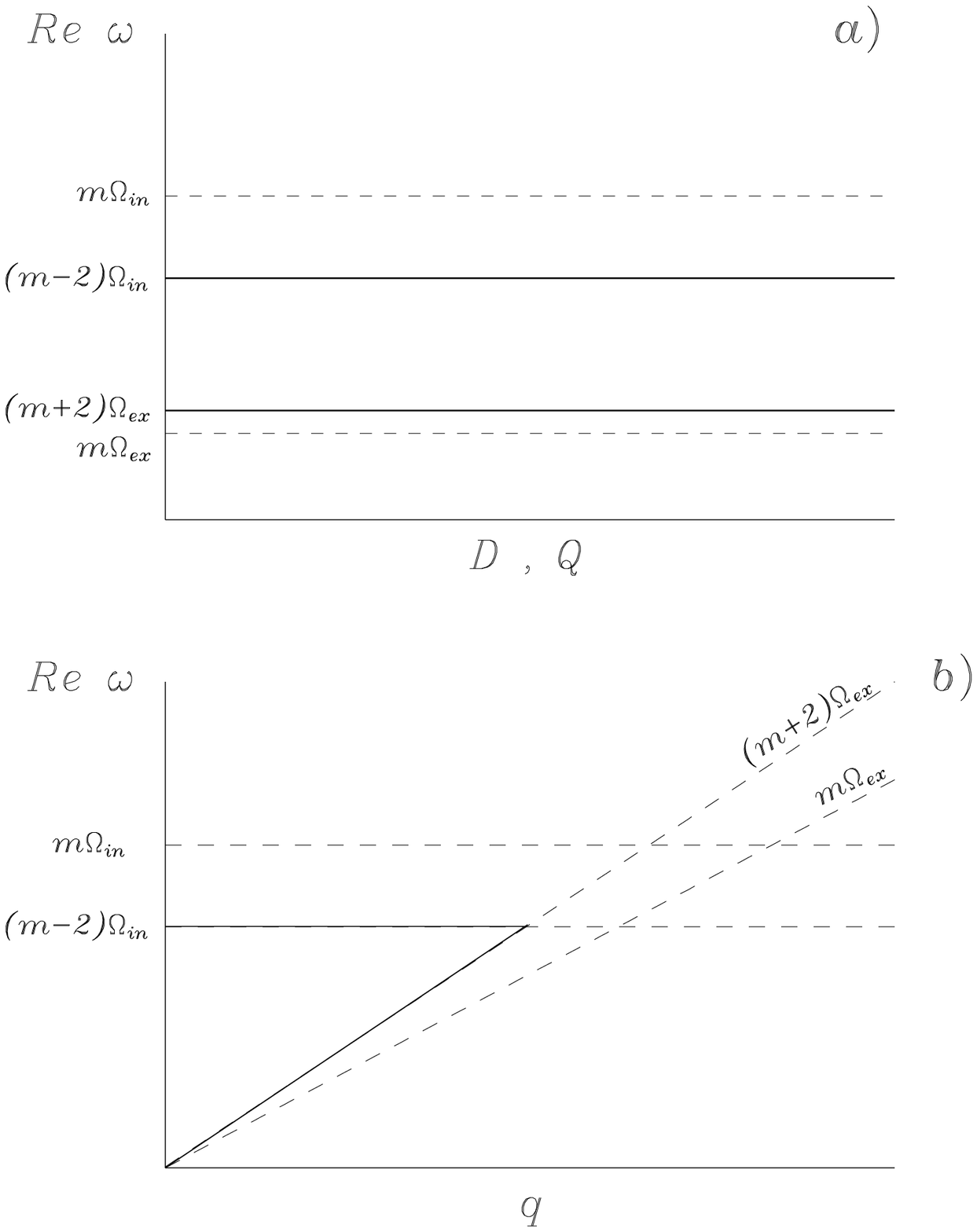}
\bigskip

{\bf Fig.3.1.} Areas of frequencies permitted for harmonics caused by
over-reflection (limited by continuous lines).

\vfil\eject

\bigskip
\epsfxsize=\hsize
\epsfbox{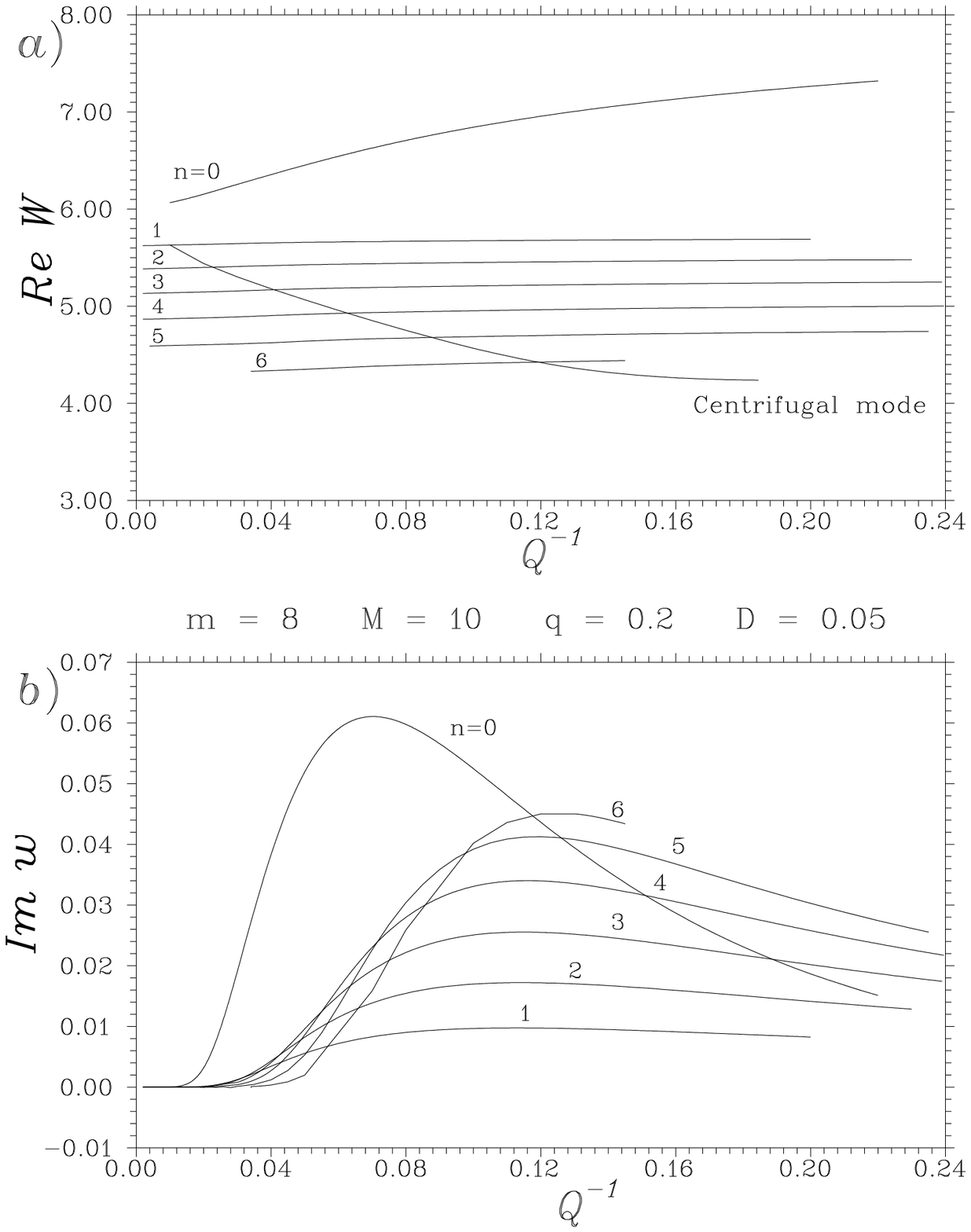}
\bigskip

{\bf Fig.3.2.} Dependence of dimensionless frequency (a) and
dimensionless growth rate (b) from relative density drop on internal
jump. The letter designations have dispersion curves of centrifugal
mode, the digits correspond to number of harmonics of gyroscopic
resonance type instability. Centrifugal instability growth rate
is not shown since it significantly exceeds GRTI growth rate.

\vfil\eject

\bigskip
\epsfxsize=0.95\hsize
\epsfbox{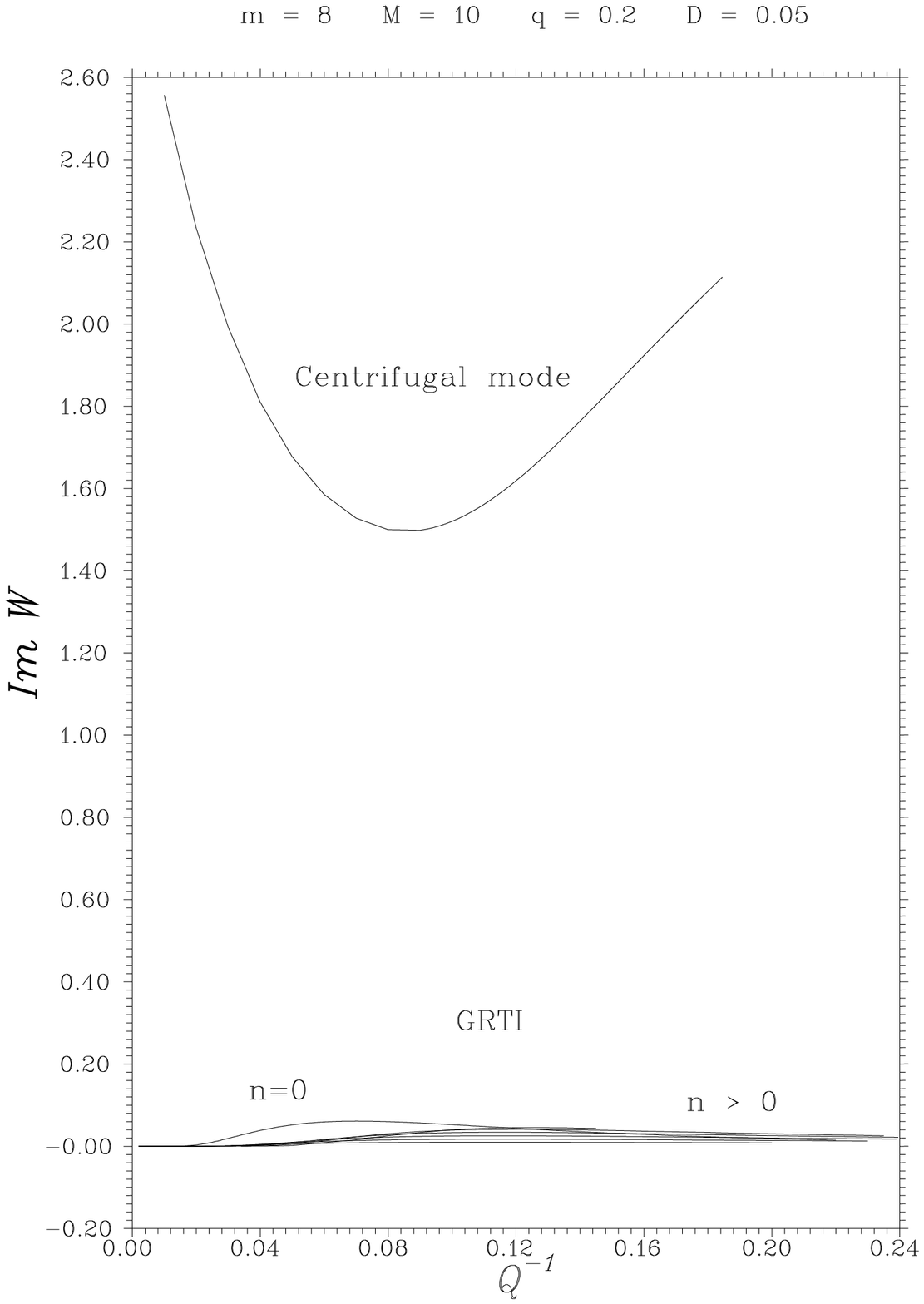}
\bigskip

{\bf Fig.3.3.} The same as on Fig. 3.2 (b) but in scale
allowing to show centrifugal instability growth rate.

\vfil\eject

\bigskip
\epsfxsize=\hsize
\epsfbox{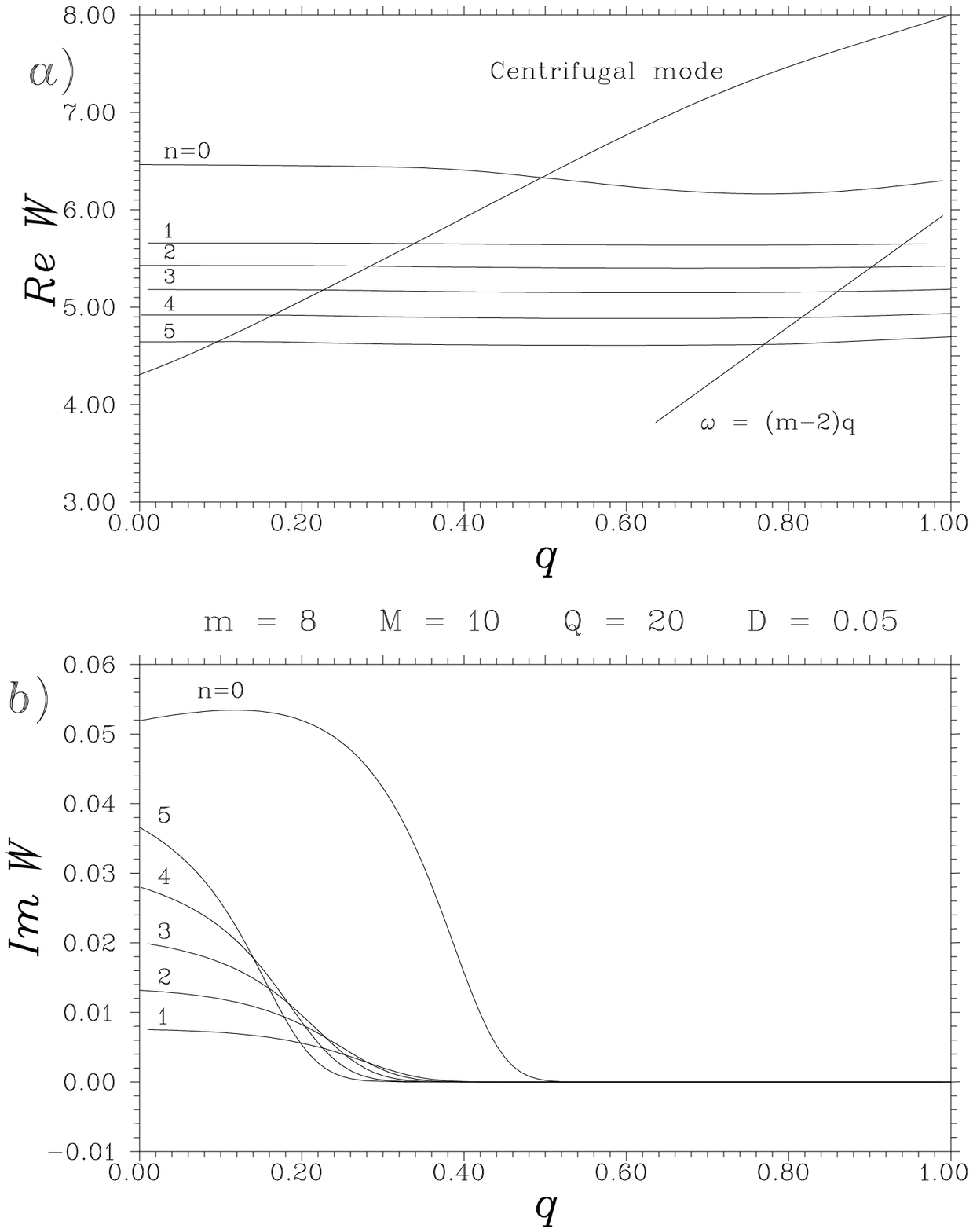}
\bigskip

{\bf Fig.3.4.} Dependence of dimensionless frequency (a) and
dimensionless growth rate (b) from angular velocity relative difference
on external jump.

\vfil\eject

\bigskip
\epsfxsize=\hsize
\epsfbox{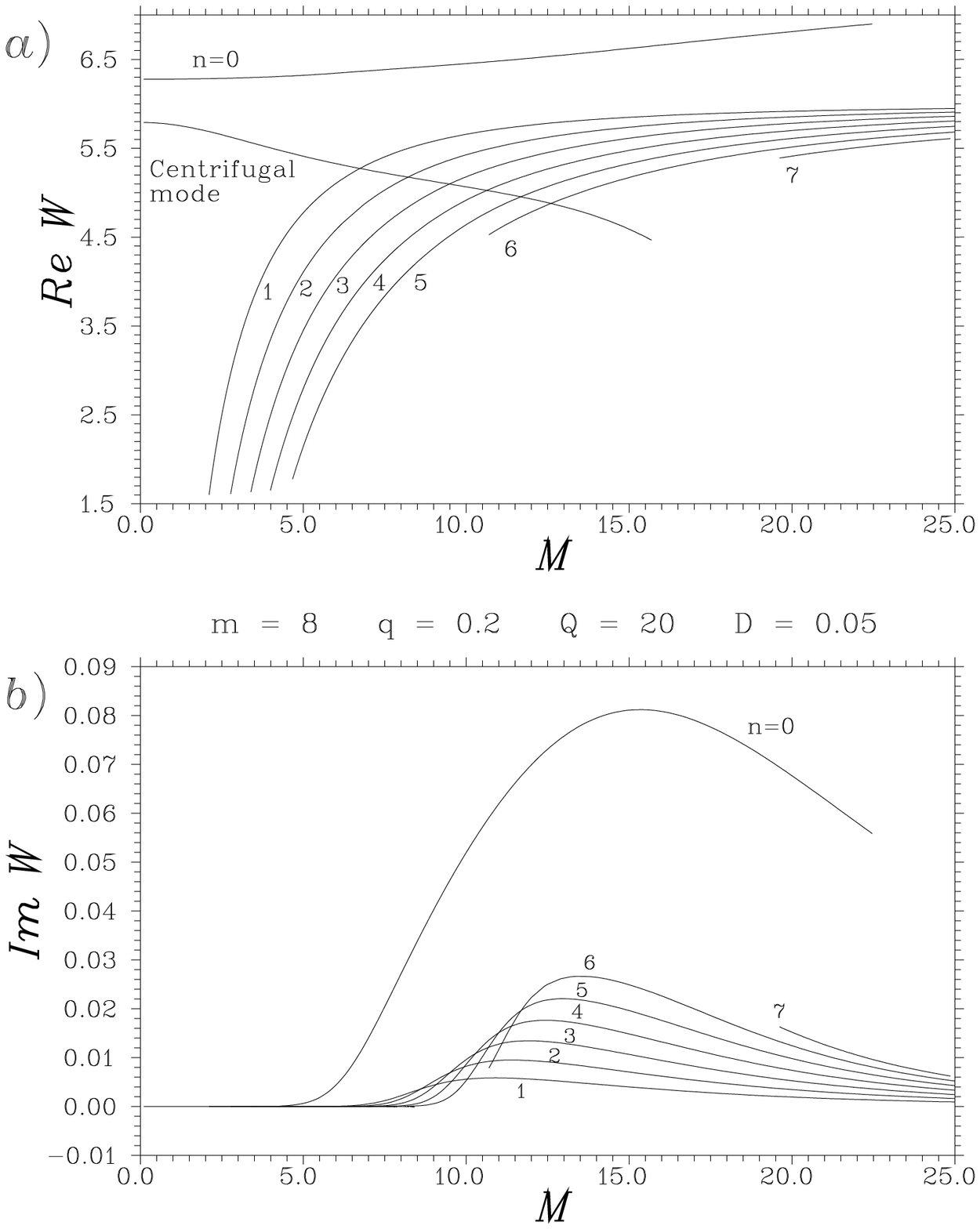}
\bigskip

{\bf Fig.3.5.} Dependence of dimensionless frequency (a) and
dimensionless growth rate (b) from Mach number.

\vfil\eject

\bigskip
\epsfxsize=\hsize
\epsfbox{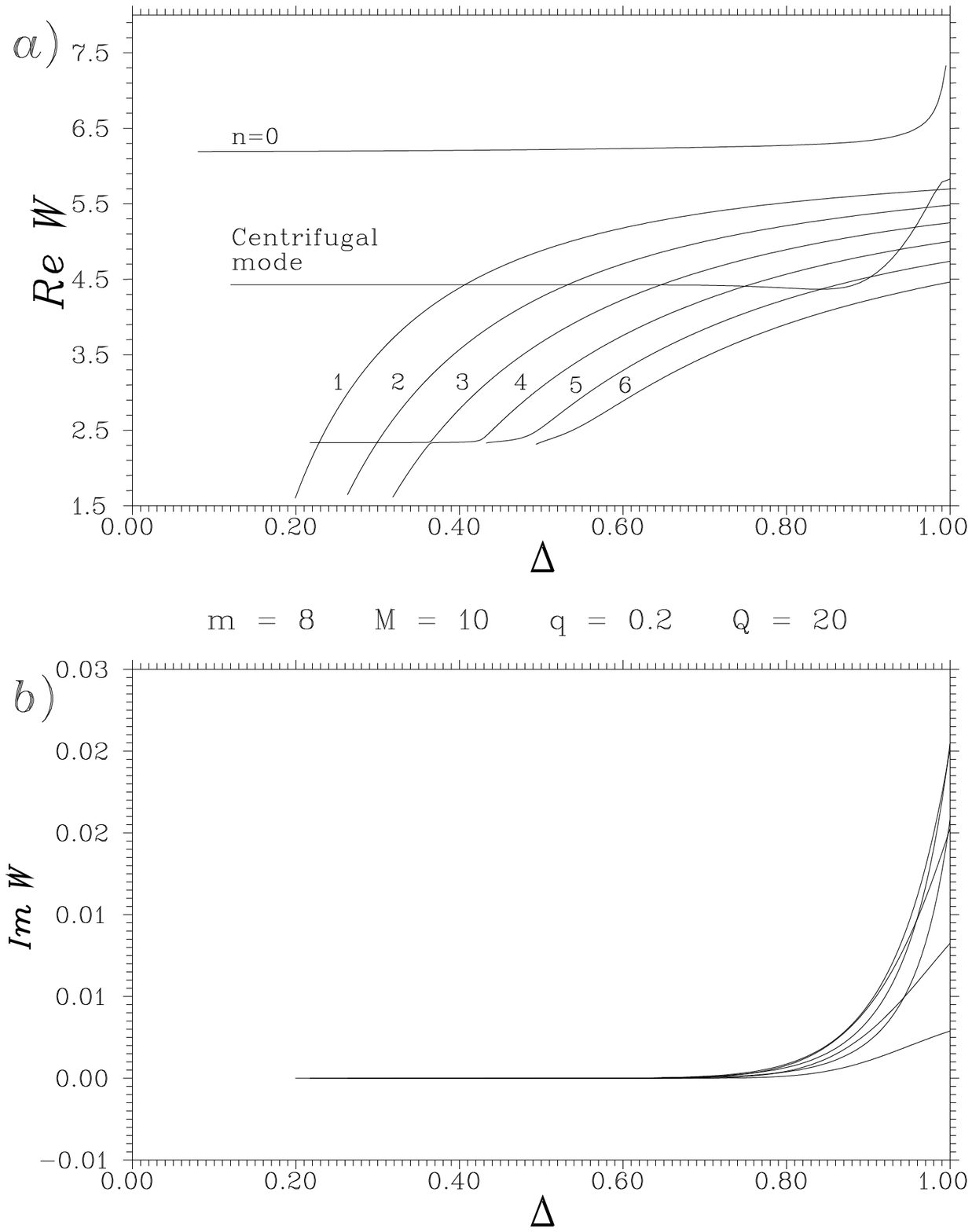}
\bigskip

{\bf Fig.3.6.} Dependence of dimensionless frequency (a) and
dimensionless growth rate (b) from the ratio of density jump radius
to velocity jump one $\Delta = R_\rho / R_\Omega = 1-D$.

\vfil\eject

\bigskip
\epsfxsize=\hsize
\epsfbox{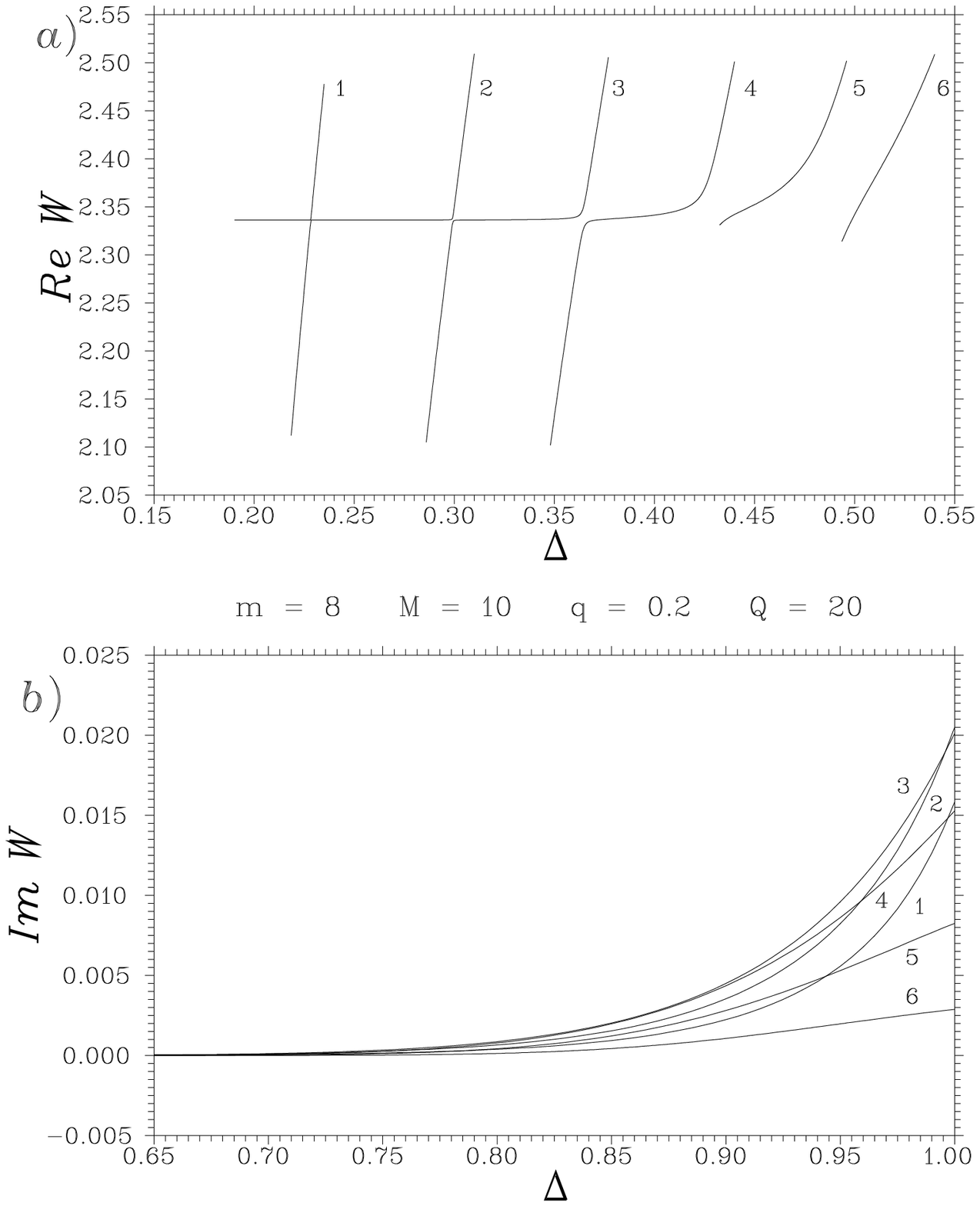}
\bigskip

{\bf Fig.3.7.} The same as on Fig. 3.6 but the most interesting
fragments are shown in larger scale.

\vfil\eject

\bigskip
\epsfxsize=\hsize
\epsfbox{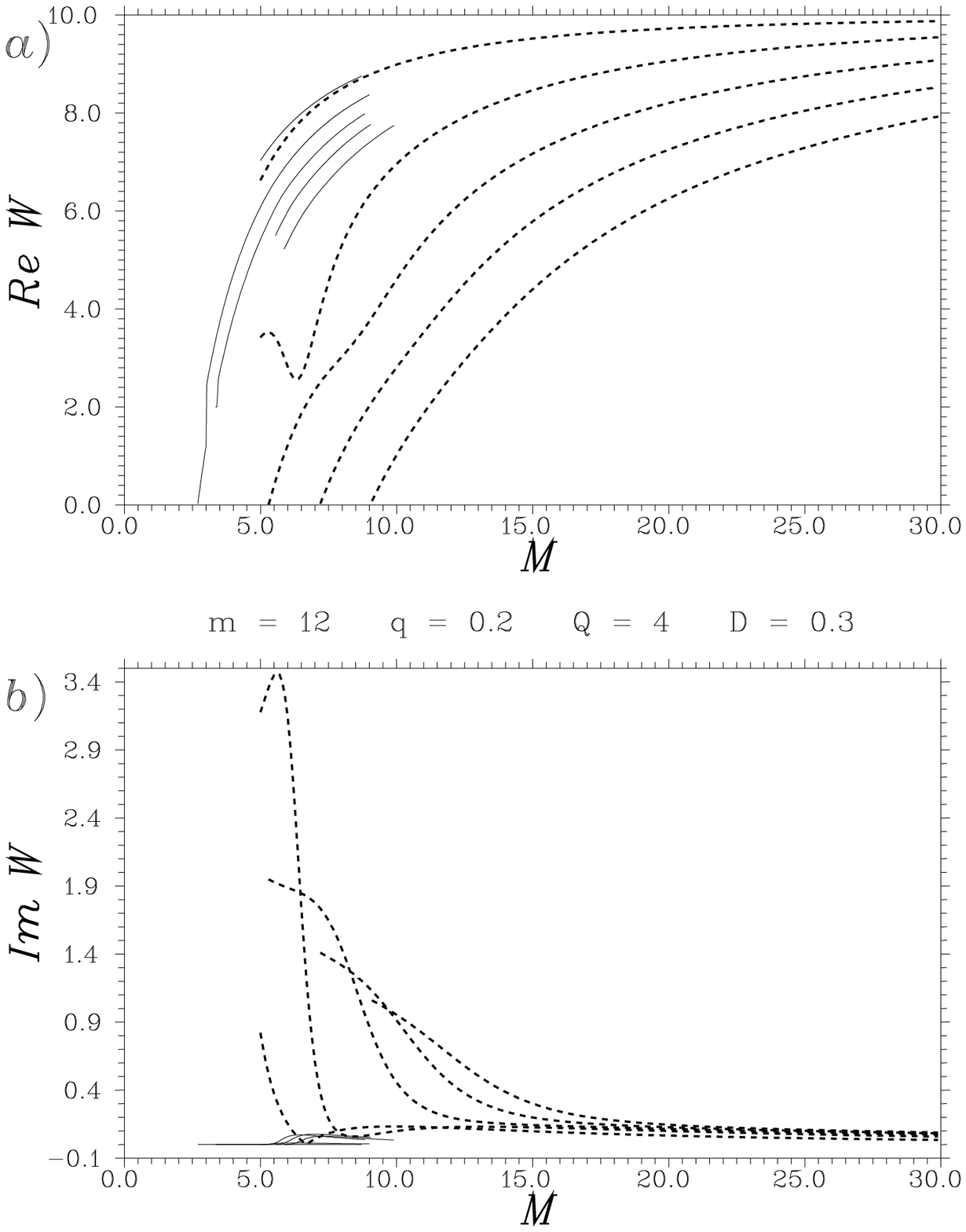}
\bigskip

{\bf Fig.3.8.} Comparative behaviour of dispersion curves by
results of the numerical solution of (5) --- continuous lines
and on asymptotics (34), (35), (37) --- dashed lines. It is visible
concurrence only qualitative.

\vfil\eject

\end